\PassOptionsToClass{pdflatex}{sn-jnl}
\documentclass[sn-nature,oneside]{sn-jnl}

\usepackage{graphicx}%
\usepackage{amsmath,amssymb}%
\usepackage{xcolor}
\usepackage{manyfoot}%

\usepackage{showyourwork}

\usepackage{siunitx}
\DeclareSIUnit\jansky{\text{Jy}}

\newcommand*\degr{\ensuremath{^\circ}}
\newcommand*\arcsec{\ensuremath{^{\prime\prime}}}   

\newcommand{\asas}{ASASSN-21qj}
\newcommand{\masyr}{mas~yr$^{-1}$}
\newcommand{\teff}{$T_\mathrm{eff}$}
\newcommand{\rsun}{$R_\odot$}

\begin{document}

\title[Planetary collision]{A planetary collision afterglow and transit of the resultant debris cloud}

\author[1]{{\fnm{Matthew} \sur{Kenworthy} }\email{kenworthy@strw.leidenuniv.nl}}
\equalcont{These authors contributed equally to this work.}

\author[2]{{\fnm{Simon} \sur{Lock} }\email{s.lock@bristol.ac.uk}}
\equalcont{These authors contributed equally to this work.}

\author[3,4]{{\fnm{Grant} \sur{Kennedy} }\email{G.Kennedy@warwick.ac.uk}}
\equalcont{These authors contributed equally to this work.}

\author[1]{{\fnm{Richelle} \sur{van Capelleveen} }\email{capelleveen@strw.leidenuniv.nl}}
\equalcont{These authors contributed equally to this work.}

\author[5]{{\fnm{Eric} \sur{Mamajek} }\email{Eric.Mamajek@jpl.nasa.gov}}
\equalcont{These authors contributed equally to this work.}

\author[6]{{\fnm{Ludmila} \sur{Carone} }\email{ludmila.carone@oeaw.ac.at}}
\equalcont{These authors contributed equally to this work.}

\author[7,8,9]{{\fnm{Franz-Josef} \sur{Hambsch} }\email{hambsch@telenet.be}}

\author[10]{{\fnm{Joseph} \sur{Masiero} }\email{jmasiero@ipac.caltech.edu}}

\author[11]{{\fnm{Amy} \sur{Mainzer} }\email{amainzer@arizona.edu}}

\author[10]{{\fnm{J. Davy} \sur{Kirkpatrick} }\email{davy@ipac.caltech.edu}}

\author[12,13]{{\fnm{Edward} \sur{Gomez} }\email{egomez@lco.global}}

\author[14]{{\fnm{Zo{\"e}} \sur{Leinhardt} }\email{Zoe.Leinhardt@bristol.ac.uk}}

\author[14]{{\fnm{Jingyao} \sur{Dou} }\email{qb20321@bristol.ac.uk}}

\author[15]{{\fnm{Pavan} \sur{Tanna} }\email{pt426@cam.ac.uk}}

\author[16]{{\fnm{Arttu} \sur{Sainio} }\email{arttu.sainio@elisanet.fi}}

\author[17]{{\fnm{Hamish} \sur{Barker} }\email{hamish.barker@gmail.com}}

\author[18]{{\fnm{St\'{e}phane} \sur{Charbonnel} }\email{stephane.charbonnel@2spot.org}}

\author[18]{{\fnm{Olivier} \sur{Garde} }\email{olivier.garde@2spot.org}}

\author[18]{{\fnm{Pascal} \sur{Le D\^{u}} }\email{pascal.ledu@2spot.org}}

\author[18]{{\fnm{Lionel} \sur{Mulato} }\email{lionel.mulato@2spot.org}}

\author[18]{{\fnm{Thomas} \sur{Petit} }\email{thomas.petit@2spot.org}}

\author[19]{{\fnm{Michael} \sur{Rizzo Smith} }\email{rizzosmith.1@osu.edu}}

\affil[1]{\orgdiv{Leiden Observatory}, \orgname{Leiden University}, \orgaddress{\street{P.O. Box 9513}, \city{Leiden}, \postcode{2300 RA}, \country{The Netherlands}}}

\affil[2]{\orgdiv{School of Earth Sciences}, \orgname{University of Bristol}, \orgaddress{\street{Queens Road}, \city{Bristol}, \postcode{BS8 1QU}, \country{UK}}}

\affil[3]{\orgdiv{Department of Physics}, \orgname{University of Warwick}, \orgaddress{\street{Gibbet Hill Road}, \city{Coventry}, \postcode{CV4 7AL}, \country{UK}}}

\affil[4]{\orgdiv{Centre for Exoplanets}, \orgname{University of Warwick}, \orgaddress{\street{Gibbet Hill Road}, \city{Coventry}, \postcode{CV4 7AL}, \country{UK}}}

\affil[5]{\orgdiv{Jet Propulsion Laboratory}, \orgname{California Institute of Technology}, \orgaddress{\street{4800 Oak Grove Drive}, \city{Pasadena}, \postcode{91109}, \state{CA}, \country{USA}}}

\affil[6]{\orgdiv{Space Research Insitute}, \orgname{Austrian Academy of Sciences}, \orgaddress{\street{Schmiedlstrasse 6}, \city{Graz}, \postcode{A-8042}, \country{Austria}}}

\affil[7]{\orgname{Vereniging Voor Sterrenkunde}, \orgaddress{\street{Oostmeers 122 C}, \city{Bruges}, \postcode{8000 Brugge}, \country{Belgium}}}

\affil[8]{\orgname{American Association of Variable Star Observers}, \orgaddress{\street{185 Alewife Brook Parkway}, \city{Cambridge}, \postcode{MA 02138}, \state{MA}, \country{USA}}}

\affil[9]{\orgname{Bundesdeutsche Arbeitsgemeinschaft f{\"u}r Ver{\"a}nderliche Sterne e. V.}, \orgaddress{\street{Munsterdamm 90}, \city{Berlin}, \postcode{D-12169}, \country{Germany}}}

\affil[10]{\orgdiv{IPAC}, \orgname{Caltech}, \orgaddress{\street{1200 E California Blvd}, \city{Pasadena}, \postcode{MC 100-22}, \state{CA}, \country{USA}}}

\affil[11]{\orgdiv{Lunar and Planetary Laboratory}, \orgname{University of Arizona}, \orgaddress{\street{1629 E. University Blvd.}, \city{Tucson}, \postcode{85721}, \state{AZ}, \country{USA}}}

\affil[12]{\orgdiv{Las Cumbres Observatory}, \orgaddress{\street{6740 Cortona Dr}, \city{Goleta}, \postcode{93117}, \state{CA}, \country{USA}}}

\affil[13]{\orgdiv{School of Physics and Astronomy}, \orgname{Cardiff University}, \orgaddress{\street{The Parade}, \city{Cardiff}, \postcode{CF24 3AA}, \country{UK}}}

\affil[14]{\orgdiv{School of Physics, H H Wills Physics Laboratory}, \orgname{University of Bristol}, \orgaddress{\street{Tyndall Avenue}, \city{Bristol}, \postcode{BS8 1TL}, \country{UK}}}

\affil[15]{\orgdiv{Institute of Astronomy}, \orgname{University of Cambridge}, \orgaddress{\street{Madingley Road}, \city{Cambridge }, \postcode{CB3 0HA}, \country{UK}}}

\affil[16]{\orgname{Independent researcher}, \orgaddress{\street{J{\"a}rvipuistonkatu 7 A 10}, \city{J{\"a}rvenp{\"a}{\"a}}, \postcode{04430}, \country{Finland}}}

\affil[17]{\orgname{Variable Stars South}, \orgaddress{\street{Rutherford Street}, \city{Nelson}, \country{New Zealand}}}

\affil[18]{\orgname{Southern Spectrocopic Project Observatory Team}, \orgaddress{\street{45, Chemin du lac}, \city{Chabons}, \postcode{38690}, \country{FRANCE}}}

\affil[19]{\orgdiv{Department of Astronomy}, \orgname{The Ohio State University}, \orgaddress{\street{140 West 18th Avenue}, \city{Columbus}, \postcode{43210}, \state{OH}, \country{USA}}}
%
%
\abstract{
Planets grow in rotating disks of dust and gas around forming stars, some of which can subsequently collide in giant impacts after the gas component is removed from the disk \cite{Williams11,Wyatt15,Hughes18}.

Monitoring programs with the warm Spitzer mission have recorded significant and rapid changes in mid-infrared output for several stars, interpreted as variations in the surface area of warm dusty material ejected by planetary-scale collisions and heated by the central star: e.g., NGC 2354–ID8 \cite{2014Sci...345.1032M,Su19}, HD 166191 \cite{Su22} and V844 Persei \cite{2021ApJ...918...71R}.
Here we report combined observations of the young ($\sim$~300~Myr), solar-like star \asas{}: an infrared brightening consistent with a blackbody temperature of 1000 K and a luminosity of 4\% of that of the star lasting for about 1000\,days, partially overlapping in time with a complex and deep wavelength-dependent optical eclipse that lasted for about 500 days.
The optical eclipse started 2.5\,years after the infrared brightening, implying an orbital period of at least that duration.
These observations are consistent with a collision between two exoplanets of several to tens of Earth masses at 2-16\,au from the central star. 
Such an impact produces a hot, highly-extended post-impact remnant with sufficient luminosity to explain the infrared observations.
Transit of the impact debris, sheared by orbital motion into a long cloud, causes the subsequent complex eclipse of the host star.
}


\maketitle


\section{Main Article}\label{sec1}

%
%
%
%
%
%
%
%
%
%

\begin{figure}
\begin{centering}
\includegraphics[width=\textwidth]{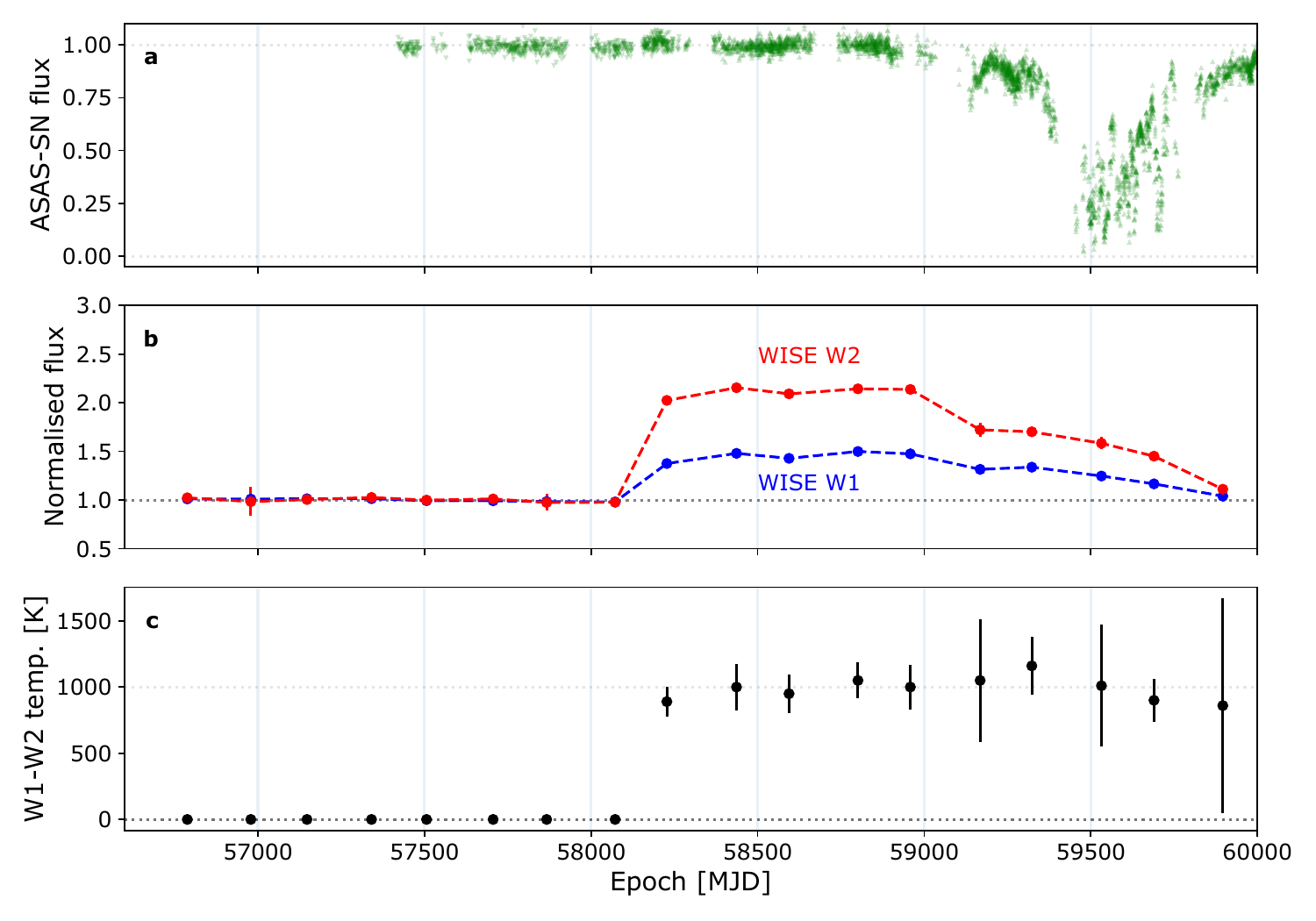}
\caption{\textbf{Optical and infrared photometry of ASASSN-21qj.}
{\bf a}, Normalized optical photometry from ASAS-SN in the $V$-band and the $g'$ band.
{\bf b},Fractional flux increase in brightness of ASASSN-21qj in both the $W1$ and $W2$ bands, for which a value of 1.0 represents the stellar contribution alone.
{\bf c},Calculated NEOWISE color temperature estimated from the photometry of the two bands.
The color temperature is plotted as zero when there is no infrared excess and is consistent with a temperature of 1000\,K while the excess is present.
Error bars are shown at $1\sigma$ confidence.
}
\label{fig:wisephot}
\script{plot_all_phot_nature_delta_flux.py}
\end{centering}
\end{figure}

The otherwise unremarkable star 2MASS J08152329-3859234 underwent a sudden optical dimming event in December 2021 \cite{RizzoSmith21,RizzoSmith22} and was assigned the identifier \asas{} by the ASAS-SN survey \citep{shappee_man_2014,kochanek_all-sky_2017}.
Here, we combine both optical (from the Las Cumbres Observatory Global Telescope, LCOGT) and infrared (from the WISE satellite) observations of \asas~for the years before and after this dimming event (Figure~\ref{fig:wisephot}). 
Optical multiband photometry shows a wavelength dependent depth consistent with extinction by sub-micron particles.
About 900 days prior to the optical dimming event, the \asas~system showed a significant brightening in the infrared, of 0.4 magnitudes at 3.8 microns ($W1$) and 0.8 magnitudes at 4.5 microns ($W2$).
Before this time the IR brightness was consistent with being purely stellar.
The increased IR fluxes indicate that in addition to the quiescent stellar flux, there was new emission at a temperature of approximately 1000\,K.
Such a remarkable combination of observations, particularly the 2.5\,year delay between the IR and optical variation, requires an explanation.

The optical and infrared light curves (Figure \ref{fig:wisephot}) provide key constraints on any proposed scenario.
The post-brightening IR fluxes in the $W1$ and $W2$ WISE passbands are consistent with emission at a blackbody temperature of $1000 \pm 100$\,K and this temperature is sustained, within error, for the remainder of our observation window, despite a decline in the total flux.
For an emitter located at the distance of ASASSN-21qj from Earth, and the observed  maximum luminosity of approximately 0.04\,$L_\star$, this implies an emitting area of 0.01\,au$^2$, equivalent to an object with a radius of 7\,$R_\odot$, or $750$\,$R_{\rm Earth}$.
If this emission was from material -- e.g., dust -- passively heated by proximity to the star, then that material must have been generated and remained within about 0.1\,au to produce the observed temperature.


One possible explanation is that we are observing two unrelated but coincidental phenomena: a warm dust generating collision within 0.1 au of the star, with a separate object transiting the star 900 days later.
Two events that are themselves very rare occurring independently in one system is, however, highly improbable.
A second explanation is that warm dust is generated close to the star and causes the optical transit, but this requires a fine-tuned configuration where the star is optically blocked by scale height variations in the resulting disk.

\begin{figure*}
\begin{centering}
\includegraphics[width=1.0\textwidth]{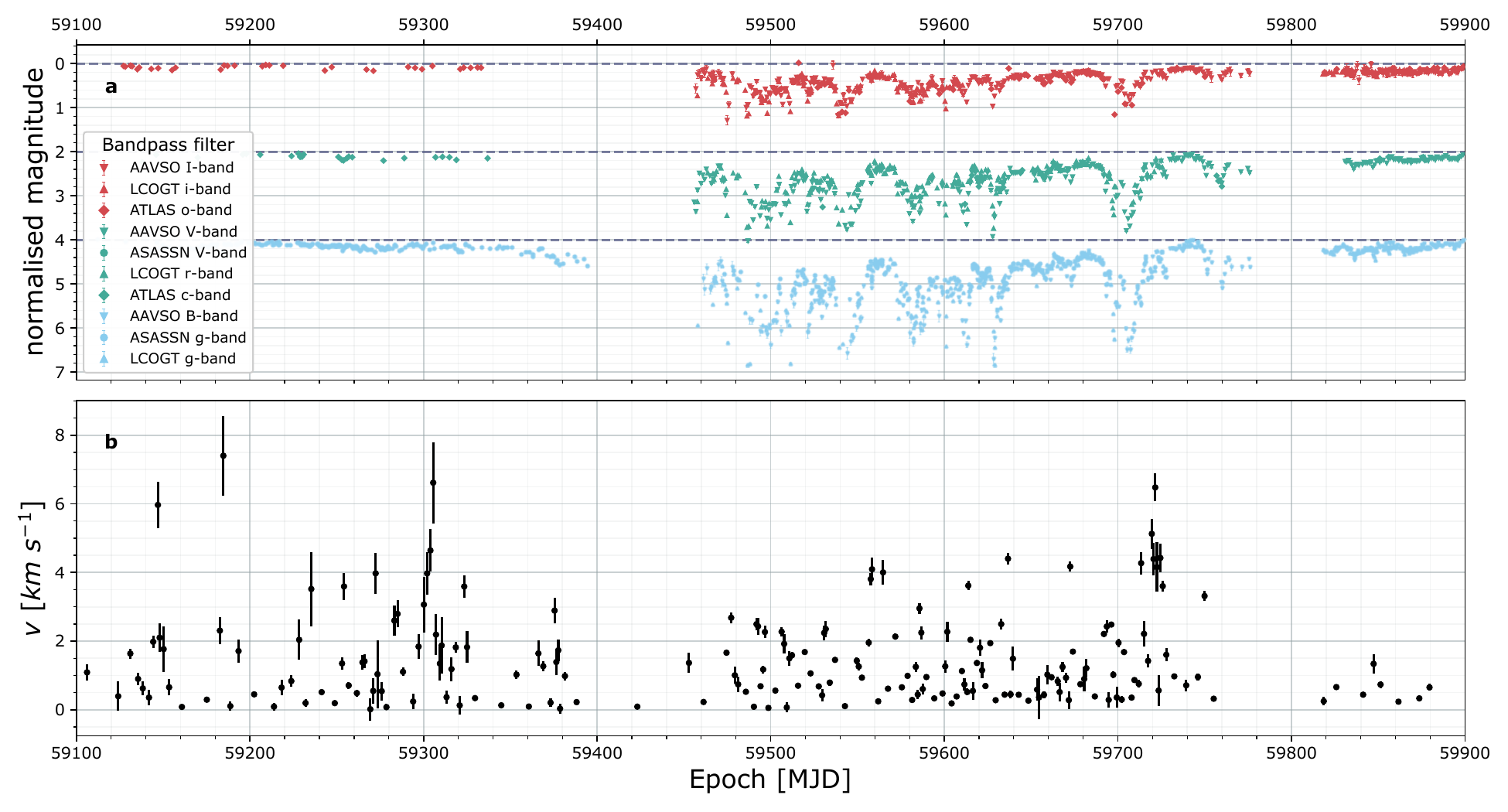}
      \caption{\textbf{The light curve of ASASSN-21qj from several different photometric surveys and the derived transverse velocities.}
      {\bf a},The eclipse depth is deeper for shorter wavelengths, indicating that the transiting material is dominated by sub-micron sized grains.
      {\bf b},Transverse velocities derived from the light-curve gradients.
      These are lower limits to the true velocity, and thus imply that the transiting material is closer to the star than 16\,au.
      Error bars are shown at $1\sigma$ confidence.
      }
        \label{fig:eclipse_overview}
\script{plot_master_lightcurve_nature_simple.py}
\end{centering}
\end{figure*}

Instead, we hypothesize that are we are observing the aftermath of a single collision between super-Earths or mini Neptunes -- a so-called giant impact -- between 2 and 16\,au from the star. 
These distances are determined, respectively, by the delay between the IR brightening and the optical eclipse (Fig. \ref{fig:wisephot}), and by gradients in the optical light curve (Fig. \ref{fig:eclipse_overview}).
In contrast to other extreme debris disk events where the star heats the dust, we propose that the infrared emission is directly from the post-impact body \cite{Lock2017,2009ApJ...704..770M}, and that impact debris produced the optical transit.
Giant impacts are a common occurrence in planet formation \cite{Schlichting2018a,DAngelo2018} and also occur during instabilities in older systems \cite{Kaib2016}; this would explain the observations with a single event of a type that are expected for systems with ages like \asas{}.

Giant impacts are one of the most energetic events planets experience.
For example, the kinetic energy of impacts between two half-Neptune-mass bodies range from $10^{33}$ to $10^{34}$~J, enough to vaporize the colliding bodies several times over.
A large fraction of this energy is dissipated in the colliding bodies and post-impact bodies are substantially melted and vaporized \cite{Nakajima2015,Lock2017,Carter2020}.
Furthermore, extreme torques exerted in impacts often produce rapidly rotating bodies \cite{Lock2017}.
Such low density and rotationally-flattened bodies can be hundreds of times larger than the pre-impact planets \cite{Lock2017} with correspondingly large radiative surfaces.

Giant impacts produce significant amounts of debris, typically around 1\% of the colliding mass, that is injected into orbit around the host star\cite{Canup2001,Lock18}.
Impact ejecta have a wide range of sizes, from sub-micron dust to planetesimals of tens to hundreds of kilometers, and often contains the most highly heated material \cite{Benz2008_Mercury_book,Leinhardt2015,Carter2020a}.
For sufficiently high impact velocities ($>1$~km~s$^{-1}$ for water ice, and $>8$~km~s$^{-1}$ for forsterite) a substantial fraction of this material is vaporised \cite{Stewart2008,Davies2020,Carter2020a}.
Shearing of droplets and cooling and condensation of vapor produces a population of small dust grains and solid spherules.
The size distribution of this fraction of the debris is uncertain, due to difficulties in modelling condensate nucleation and breakup, but previous work suggests that debris could range in size from sub-micron to decimeters \cite{Benz2008_Mercury_book,Johnson2015}.
The wavelength-dependent eclipse suggests that the optical depth of the transiting dust cloud is dominated by sub-micron grains, consistent with these previous estimates.
While the fading of the excess infrared flux occurs within 100 days of the start of the optical transit, we consider the timing to be coincidental because there is no clear correspondence between the light curves, for example no  change in the (six month cadence) IR flux when the star dims in the optical wavelengths just before MJD 59500.

The radiative flux from post-impact bodies has not been explored in depth.
Computational resource limitations make resolving the low-density outer regions and photosphere of post-impact bodies extremely challenging.
However, preliminary simulations of impacts between super-Earth and mini-Neptunes place an approximate lower limit on the extent of post-impact bodies and show that post-impact bodies can extend to hundreds of Earth radii.
Such an object radiating at $\sim$1000~K would produce a flux comparable to the 0.04~$L_*$ inferred from our observations. 

\begin{figure}
\begin{centering}
\includegraphics[width=\columnwidth]{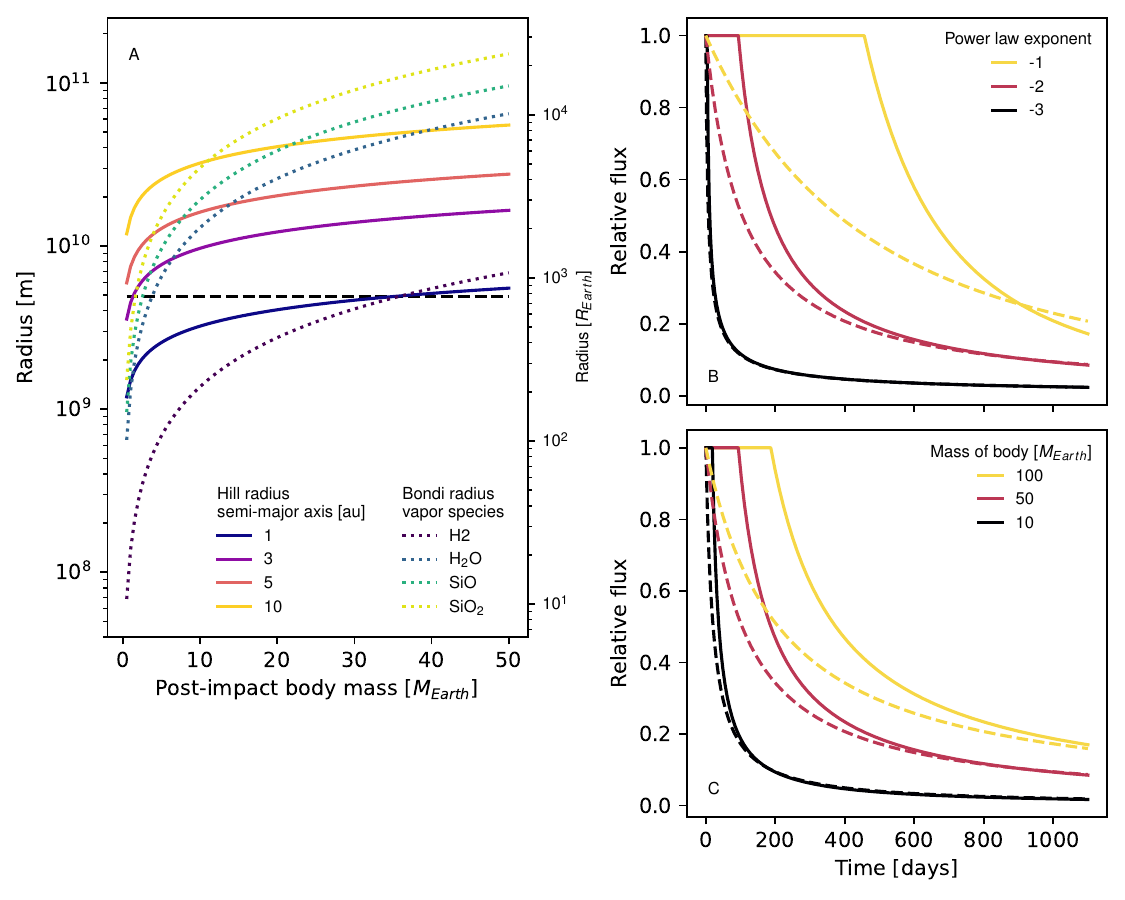}
\caption{\textbf{The size and temporal evolution of a post-impact body.}
A post-impact body of a few to tens of Earth masses could be large enough to explain the observed increase in thermal emission from ASASSN-21qj and the subsequent infrared fluxes.
{\bf a},The Hill radii for bodies at different semi-major axes (solid colored lines) and the Bondi radii (dashed lines) assuming different compositions for the vapor of the post-impact body.
The horizontal black dashed line shows the radius required to explain the observed flux at the inferred emission temperature of 1000~K. 
{\bf b,c},The change in excess flux owing to a post-impact body with time for a simple model of cooling of a post-impact body for different power-law surface density profiles ({\bf b}; equation~\ref{eqn:sigma}) and for different mass bodies ({\bf c}).
Each example had an initial radius of 7\,$R_\odot$, a radiative temperature of 1000~K, and a corresponding initial flux of 0.04\,$L_\star$, as estimated for the observed infrared emitter.
The solid lines are profiles that can have non-zero initial density at the initial emitting radius, and the dashed lines are for ones that are forced to have zero density at the initial emitting radius.
When not stated, the mass of the post-impact body was 50~$M_{\rm Earth}$, the power law exponent is $-2$, and the size of the central region was 10~$R_{\rm Earth}$.
}
\label{fig:Hill_Bondi_R}
\script{plot_hill_bondi_radii_three_panels.py}

\end{centering}
\end{figure}

Independent of impact simulations, there are fundamental limits on the size of post-impact bodies from the Hill and Bondi radii.
Any post-impact structure must lie within the Hill sphere, the distance within which the gravity of an object dominates over that of the star.
Furthermore, only vapor within the Bondi radius would be bound to the post-impact body.
Figure~\ref{fig:Hill_Bondi_R}A shows the Hill radii at different distances from ASASSN-21qj (solid lines) and the Bondi radii for example gas species (dotted lines).
Beyond 2.4\,au, the Hill radii of greater than Earth-mass bodies are large enough to accommodate a post-impact body capable of producing the required IR flux ($\sim7R_*$, dotted line). 
Heavier gases (H$_2$O, SiO, and SiO$_2$) are also bound to bodies of more than a few Earth masses.
A post-impact body of a few Earth masses can hence theoretically produce the observed IR emission, with a photosphere that is dust and/or vapor.

A key line of evidence for direct detection of a post-impact body is the constant emission temperature.
It is argued \cite{Lock18} that rocky post-impact bodies become optically thin at low pressures where radiative loss drives rapid condensation of the rock vapor.
The emission temperature is then set by the liquid-vapor phase boundary and is constant until the post-impact body almost fully condenses \cite{Lock18,Caracas2023}.
The dew point (onset of condensation) and bubble point (onset of substantial vaporization) of material with bulk silicate Earth composition are similar \cite[$\sim2300$~K, within $\sim 100$~K;][]{Lock18,Fegley2023_BSE_cond} at the low pressures of the photosphere of a post-impact body.
However, even small amounts of water (10$^{-3}$ mole fraction) may lower the bubble and dew points for silicates by $\sim$100~K \cite[][]{Fegley2023_BSE_cond,Lock18}. 
The emission temperature of bodies produced by collisions of proto-planets composed of rock and ices/volatiles could be buffered at $\sim$1000~K during early evolution.


The temporal variation of flux from a post-impact body is controlled by evolution of the bodies' size which is governed by a number of competing factors \cite{Lock18,Lock2020}, including radiative energy loss, viscous spreading, and mass and angular momentum transfer by condensates.
For silicate-dominated post-impact bodies, the high emission temperature means that radiative cooling dominates, and the post-impact body contracts rapidly, fully condensing over years to thousands of years \cite{Lock18,Lock2020}.
The body we observed has a much lower emission temperature, and contracted substantially more slowly.
Figure~\ref{fig:Hill_Bondi_R}~B and C show evolution of conceptual post-impact bodies in the limiting case that radiative cooling dominates.
If sufficient mass is injected into the outer regions (i.e., less negative power law exponents) the observed flux can remain constant for an initial period and then decays over the order of hundreds of days, in agreement with infrared observations.
Further work is required to understand the structure and evolution of bodies produced by impacts between super-Earths and mini-Neptunes of different compositions and identify temporal flux variations consistent with our observations.

\clearpage

\setcounter{figure}{0}    

\renewcommand\figurename{Extended Data Fig.}
\renewcommand\tablename{Extended Data Table}%


\section{Methods}\label{sec:methods}

The stellar properties of \asas\ (Gaia DR3 5539970601632026752 = 2MASS~J08152329-3859234) are listed in Extended Data Table~\ref{tab:Stellarprop}, showing that \asas\ is consistent with being a G2 type dwarf star.
Where necessary we assume a stellar mass equal to the Sun.
\asas\  has a neighbor (Gaia DR3 5539970597334497024 = 2MASS~J08152298-3859244) which is a visual double.
Based on the Gaia DR3 mean ICRS position for epoch 2016.0, the visual companion lies at a separation $\rho$ = $3738.243\pm0.062$ mas and at position angle $\theta$ = $249^{\circ}.977$.
Their parallaxes ($\varpi$ = $1.7631\pm0.0112$ mas vs. $1.4711\pm0.0523$ mas) differ by 5.5$\sigma$ and proper motions ($\mu_{\alpha} = -9.692\pm0.012$, $\mu_{\delta} = 7.349\pm0.012$ \masyr\, vs. $\mu_{\alpha} = -0.114\pm0.055$, $\mu_{\delta} = 6.419\pm0.053$ \masyr) differ by a factor of 2.
The large differences in distance and proper motions suggests that these stars are not associated.

\begin{figure}[h!]
    \centering
\includegraphics[width=0.8\textwidth]{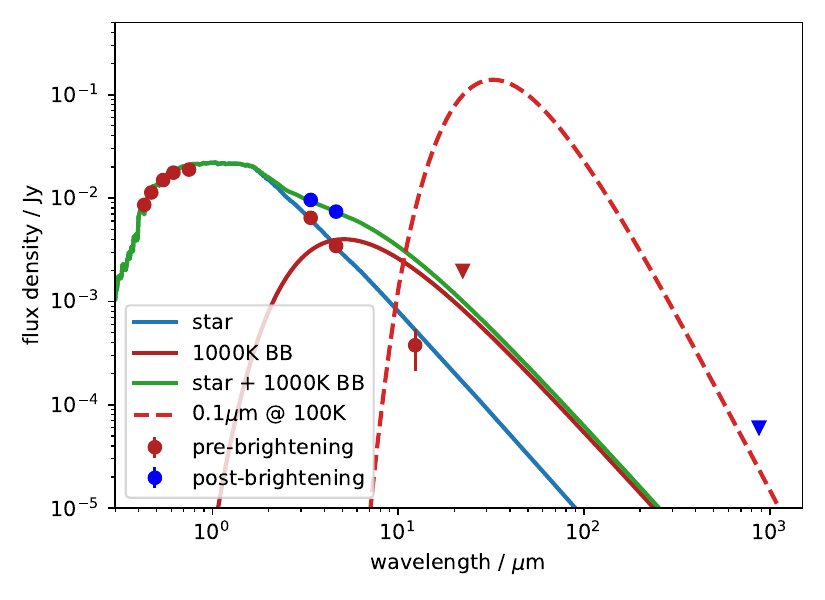}
    \caption{\textbf{Spectrum of \asas{}~components.}
    The red symbols show the optical and pre-brightnening WISE infrared photometry and the blue symbols show the post-brightnening WISE and ALMA fluxes.
    Stellar and 1000\,K components consistent with the pre- and post-brightening fluxes are shown.
    The dashed line shows an estimated cool component spectrum for $0.1\mu m$-sized grains associated with the transiting dust cloud.
    Downward triangles are upper limits.
    Error bars are shown at $1\sigma$ confidence.
}
    \label{fig:sed}
   \script{fvsr.py}
\end{figure}

The stellar photospheric flux was estimated by fitting stellar models to GAIA, APASS, and DENIS and WISE optical/near-IR photometry (the 2MASS $J$ photometry is an upper limit, and $H$ and $K_s$ are flagged as contaminated).
Extended Data Figure \ref{fig:sed} shows the resulting models (the dashed line is discussed below).
The ALLWISE photometry ($\sim$2010) is consistent with the NEOWISE photometry pre-brightening nearly ten years later.

A fit using the method of \cite{2019MNRAS.488.3588Y} with GAIA and DENIS photometry finds that the WISE W1/2 fluxes are about 20\% too high, but better agreement is found with APASS photometry.
The difference is explained by the fact that WISE and APASS have lower spatial resolution and include the flux of the $\sim$2\,mag fainter visual double to the West of \asas{} (which is visible in 2MASS).
The best fit stellar effective temperature is $5560 \pm 100$\,K.
We do not find that reddening is needed for these models; while there is relatively little photometry with which to strongly constrain both $T_{\rm eff}$ and $A_V$, the $T_{\rm eff}$ from GAIA DR3 is $5760 \pm 10$\,K, which suggests that the conclusion of little reddening is valid.
Primarily, we conclude that there is no evidence that \asas{} showed evidence for an IR excess before the brightening seen by NEOWISE.

To estimate the IR excess properties we fit the same models, but now including the first five post-brightening NEOWISE data points.
We correct for the nearby source by using the GAIA photometry for the star, and the post/pre-brightening difference for the WISE excess flux.
This fit yields the fractional luminosity $L_{\rm dust}/L_\star = 0.04 \pm 0.005$ and a dust temperature $950 \pm 30$\,K.
To estimate the dust temperature as a function of time we simply subtract the median pre-brightening W1/2 fluxes, with the RMS of these values as the uncertainty (see Figure \ref{fig:wisephot}).
The temperature uncertainty increases as the excess fades; the excess flux uncertainty depends on both the observed flux and the stellar flux, and as the excess decreases the stellar uncertainty (which is constant) becomes an increasingly large fraction of the excess.
These fluxes therefore exclude the contaminating flux from the nearby object, and are consistent with the SED-derived dust temperature.

Converting dust temperatures to stellocentric radii is uncertain because dust temperature depends on grain size.
Typically the radius derived under the assumption of blackbody emission is an underestimate by a factor of up to five \citep{2013MNRAS.428.1263B,2015MNRAS.454.3207P}.
Thus, we conclude that the dust location might in the most extreme case be as far as 1\,au, but not sufficiently far to explain the 900\,day delay between the WISE and optical flux variations. which requires a distance of at least 2\,au.

\begin{table}
    \centering
    \caption{Properties of \asas}
    \begin{tabular}{@{}lc@{}}
    \hline\hline
Property                               & Value                    \\
        \hline
         $\alpha_{ICRS}$, {[}hh mm ss{]}  & 08:15:23.30\footnotemark[1]  \\
         $\delta_{ICRS}$, {[}dd mm ss{]}  & -38:59:23.3\footnotemark[1]  \\
         $\mu_{\alpha}$ {[}mas yr$^{-1}${]}     & $-9.692\pm0.012$\footnotemark[1]   \\
         $\mu_{\delta}$ {[}mas yr$^{-1}${]}     & $7.349\pm0.012$\footnotemark[1]  \\
         $\varpi$ {[}mas{]}                     & $1.763\pm0.011$\footnotemark[1]   \\
         RV {[}km s$^{-1}${]}                   & $25.8\pm3$\footnotemark[1] \\
         Distance {[}pc{]}                      & $567.2^{+6.7}_{-5.9}$\footnotemark[2] \\ 
        \hline
         $G$ {[}mag{]}                          & $13.371\pm 0.003$\footnotemark[1]  \\
         $G_{BP}$ {[}mag{]}                     & $13.697\pm 0.003$\footnotemark[1]    \\
         $G_{RP}$ {[}mag{]}                     & $12.882\pm 0.004$\footnotemark[1]   \\
         $G_{BP}-G_{RP}$ {[}mag{]}              & $0.815\pm 0.005$\footnotemark[1]         \\
         $J$ {[}mag{]}                          & $>12.07$\footnotemark[3]   \\
         $H$ {[}mag{]}                          & $12.03\pm0.04$\footnotemark[3]    \\
         $K_s$ {[}mag{]}                          & $11.99\pm0.04$\footnotemark[3]    \\
         $B$ (AB) {[}mag{]}                     & $14.16\pm0.06$\footnotemark[4]     \\
         $V$ (AB) {[}mag{]}                     & $13.48\pm0.03$\footnotemark[4]   \\
         $g'$ (AB) {[}mag{]}                     & $13.77\pm0.03$\footnotemark[4]  \\
         $r'$ (AB) {[}mag{]}                     & $13.29\pm0.03$\footnotemark[4]     \\
         $i'$ (AB) {[}mag{]}                     & $13.20\pm0.08$\footnotemark[4]   \\
        \hline
         \teff{} [K]                            & $5760\pm10$\footnotemark[1]  \\
         {[}Fe/H{]} [dex]                       & $-0.23\pm0.01$\footnotemark[1] \\
         log\,$g$ [log$_{10}$ cm\,s$^{-2}$]     & $4.339\pm0.005$\footnotemark[1]  \\
         $R_*$ [\rsun{}]                        & $1.04$\footnotemark[5] \\
         log($L_{bol}/L_{\odot}$) [dex]         & $0.033$\footnotemark[5] \\
         $m_{bol}$ [mag]                        & $13.49\pm0.02$\footnotemark[6] \\
         $E(G_{BP}-G_{RP})$ {[}mag{]}           & $0.01 \pm 0.005$\footnotemark[6]    \\
         Age [Myr]                              & $300\pm92$\footnotemark[7] \\
         
        \hline
    \end{tabular}
    \footnotetext[1]{Gaia DR3 \cite{Brown2023}, coordinates are J2000 at epoch 2000.0.}
    \footnotetext[2]{\cite{BailerJones21}}
    \footnotetext[3]{2MASS \cite{Cutri03}, $J$ is an upper limit and $H$ and $K_s$ are flagged as contaminated.}
    \footnotetext[4]{APASS \cite{2015AAS...22533616H}, this photometry includes a nearby star.}
    \footnotetext[5]{Estimated from an SED fit fixed to Gaia properties.}
    \footnotetext[6]{Estimated using mean stellar properties \cite{2013ApJS..208....9P}.}
    \footnotetext[7]{Estimated using rotation period.}

\label{tab:Stellarprop}
\end{table}

\subsection*{Observations}\label{sec:obs}

The beginning of the eclipse was announced \cite{RizzoSmith21} by the ASAS-SN survey, which triggered several observing campaigns at optical wavelengths and an ALMA observation at Band 7 (program \texttt{2019.A.00040.S}).

\begin{table}
    \centering
    \caption{Photometric observations of ASASSN-21qj. The number of points are listed per survey and filter. %
    This is the count after initial rejection of photometric points with significantly large error bars.
    Further photometric points may have been rejected in the different analysis steps.
    The ROAD observations constitute the majority of observations from the AAVSO data sets.}
    \begin{tabular}{@{}lcc@{}}
    \hline\hline
Survey name  & Filter & Number of points                    \\
        \hline
ASASSN   &  $V$   & 758 \\
   &  $g'$   & 3225 \\
        \hline
ATLAS & $c$ & 161 \\
 & $o$ & 677 \\
        \hline
AAVSO   &  $B$   & 729 \\    
   &  $I$   & 728 \\    
   &  $V$   & 722 \\    
        \hline
LCOGT   &  $g$   & 275 \\    
   &  $r$   & 224 \\    
   &  $i$   & 168 \\    
        \hline
NEOWISE & $W1$ & 18 \\
 & $W2$ & 18 \\
        \hline
    \end{tabular}
 
\label{tab:photometry}
\end{table}

%
%
The All Sky Automated Survey \cite[ASAS; ][]{pojmanski_all_1997, asas_2005, asas_2018} is a survey consisting of two observing stations - one in Las Campanas, Chile and the other on Maui, Hawaii. 
Each observatory is equipped with two CCD cameras using V and I filters and commercial f $ = 200$ mm, D $= 100$ mm lenses, although both larger (D $=250$ mm) and smaller (50-72 mm) lenses were used at earlier times.
The majority of the data are taken with a pixel scale of $\approx$ 15\arcsec{}.
ASAS splits the sky into 709 partially overlapping 9\degr{} $\times$ 9\degr{} fields, taking on average 150 3-minute exposures per night, leading to a variable cadence of 0-2 frames per night.
Depending on the equipment used and the mode of operation, the ASAS limiting magnitude varied between 13.5 and 15.5 mag in V, and the saturation limit was 5.5 to 7.5 mag. 
Precision is around 0.01-0.02 mag for bright stars and below 0.3 mag for the fainter ones. 
ASAS photometry is calibrated against the Tycho catalog, and its accuracy is limited to 0.05 mag for bright, non-blended stars.


The All Sky Automated Survey for Supernovae \cite[ASAS-SN; ][]{shappee_man_2014,kochanek_all-sky_2017} consists of five stations around the globe, with each station hosting four telescopes with a shared mount.
The telescopes consist of a 14-cm aperture telephoto lens with a field of view of approximately 4.5\degr{}$\times$4.5\degr{} and an 8.0\arcsec{} pixel scale.
Two of the original stations (one in Hawaii and one in Chile) were initially fitted with $V$ band filters, but now these and all the other stations (Texas, South Africa and a second in Chile) observe with $g'$ band filters down to 18 mag.


The Remote Observatory Atacama Desert \cite[ROAD; ][]{Hambsch12} is a fully automated telescope located in Chile that obtains nightly photometry in Astrodon B, V and I bands for a wide range of astronomical projects.
It consists of a 40-cm $f/6.8$ Optimized Dall-Kirkham and uses a Finger Lakes Instruments camera with a 4k$\times$4k array with pixels of $9\mu m$ in size.
Data are reduced using a custom pipeline and then published on the AAVSO website.


Las Cumbres Observatory Global Telescope (LCOGT) is a network of 25 fully robotic operated telescopes distributed over 7 sites located all around the globe.
These telescopes are designed to observe transient astronomical events at optical and near-infrared wavelengths.
LCOGT provides a large variety of filter options, but the data we collected are in SDSS $g'$, $r'$ and $i'$ bands.
All data is automatically processed and calibrated by the BANZAI pipeline.
The visual companion caused complications in BANZAIs automatic aperture extraction routine; sometimes correct apertures were extracted for both \asas\ and the nearby star, and sometimes both sources were extracted in one large aperture, often with an offset from the true center of \asas.
To correct this, the last two stages of the BANZAI routine, aperture extraction and photometry calibration, were modified for this specific situation. 
The calibrated magnitudes of all sources in the frames are computed using the default BANZAI photometry calibration routine.


ATLAS is a project that searches for near earth asteroids down to a magnitude of 19 \cite{Tonry18}.
Two filters were obtained, the $o$ (orange) and $c$ (cyan) filters respectively.
The data consists of two to four photometric points observed each night when conditions permitted.
Photometry with large errors was rejected in a first pass, then the remaining observations during a night were averaged and an error based on the r.m.s. of these nightly points was calculated.
The photometry covers the time period where the collision event occurred. 




The Transiting Exoplanet Survey Satellite \cite[TESS; ][]{2015JATIS...1a4003R} is a satellite designed to survey for transiting exoplanets among the brightest and nearest stars over most of the sky.
The TESS satellite orbits the Earth every 13.7 days on a highly elliptical orbit, scanning a sector of the sky spanning 24\degr $\times$ 96\degr\ for a total of two orbits, before moving on to the next sector. 
It captures images at a 2 second (used for guiding), 20 seconds (for 1000 bright asteroseismology targets), 120 seconds (for 200 000 stars that are likely planet hosts) and 30 minute (full frame image) cadences.
The instrument consists of 4 CCDs each with a field of view of 24\degr$\times$24\degr, with a wide band-pass filter from 600-1000 nm (similar to the $I_C$ band) and provides high precision ($\approx$milli-mag) light curves for stars down to about 14\,mag ($I_C$).


The Near-Earth Object Wide-field Infrared Survey Explorer (NEOWISE) is a space-based infrared telescope that has been surveying the sky since 2013 at $3.4$ and $4.6~\mu$m.
NEOWISE orbits near the Earth's day-night terminator, scanning rings of the sky near $\sim90^\circ$ Solar elongation, and obtains a sequence of observations of a given region of sky every six months.
The two wavelength channels are obtained simultaneously through a beamsplitter, allowing for color information to be extracted for each source detected in both bands.
Detailed descriptions of NEOWISE operations and early results from the Reactivation mission \cite{mainzer14neowise} and the standard data processing and data characteristics \cite{cutri15} are available.
A single data point for each epoch and wavelength is calculated by taking the weighted average of the individual NEOWISE measurements.


The ALMA data from programme \texttt{2019.A.00040.S} were downloaded and processed through to a measurement set with CASA \cite{2007ASPC..376..127M}.
These observations were taken 2021 September 28 (MJD=59485) and used band 7, with a mean wavelength of 880\,$\mu$m.
No source is visible in the default archive products, and we also detect no source at the expected location in CLEAN images.
The archive products report an RMS of 17\,$\mu$Jy, and we measure an RMS of 20\,$\mu$Jy in a naturally weighted image.
We therefore consider these results as an upper limit of 60\,$\mu$Jy.
In terms of the infrared excess visible in the mid-IR with WISE this upper limit is not at all constraining, but does set limits on emission from cooler dust (Figure~\ref{fig:sed}).

\subsection*{Light curves}

In this section we consider the light curves; first the implications of the TESS data for the stellar age, and then the implications of the ground-based optical light curves for transverse velocity of the occulting material and dust grain sizes.

\subsection*{Stellar rotation and age}

\begin{figure}
   \begin{centering}
\includegraphics[width=\textwidth]{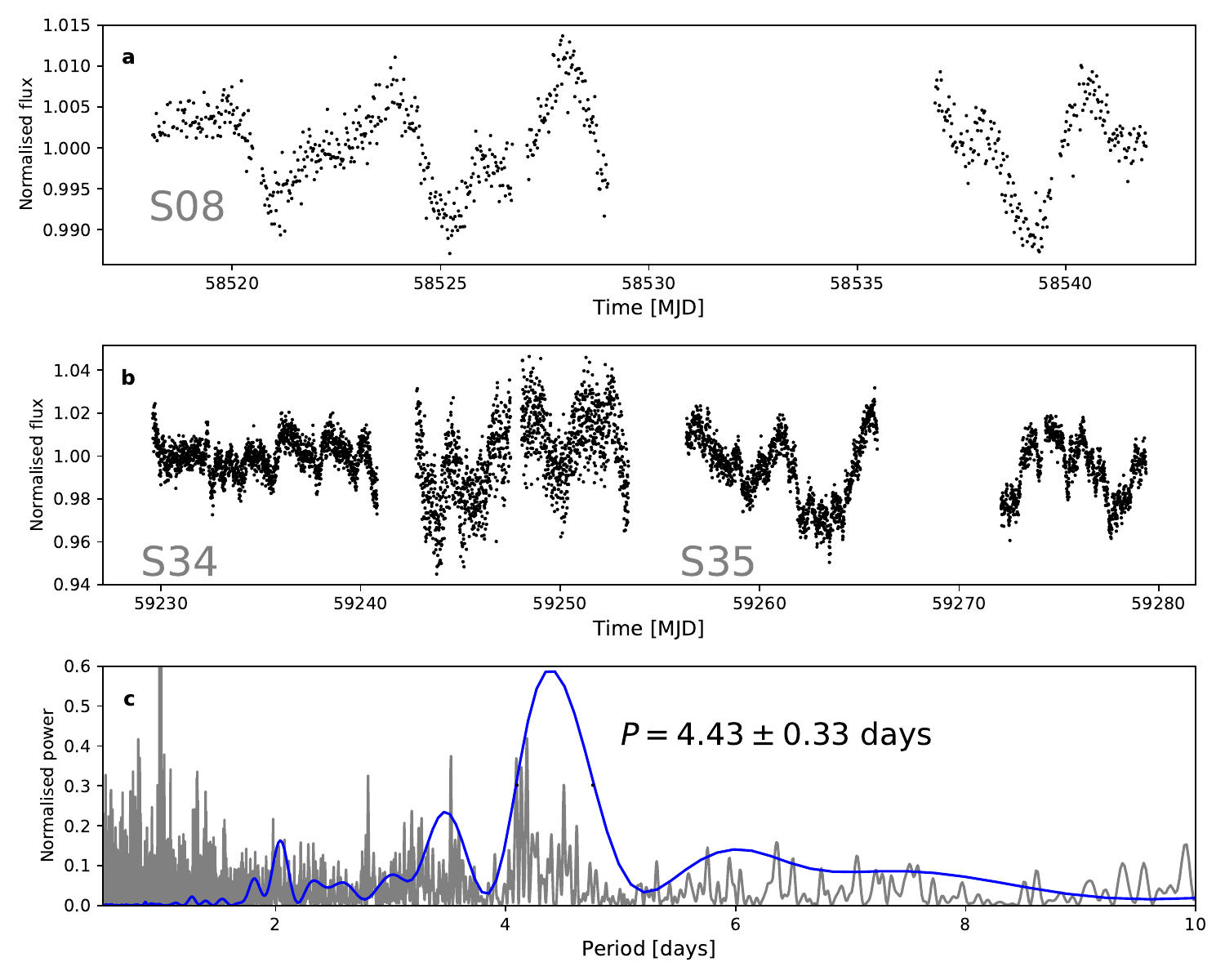}
      \caption{\textbf{The light curve of ASASSN-21qj from TESS and the periodogram of TESS and ASAS-SN photometry.}
      {\bf a,b},Photometry of ASASSN-21qj from three sectors of TESS.
      {\bf c},Lomb-Scargle analysis of photometry from TESS (colored blue) and from ASAS-SN V band data from MJD 57420 to MJD 58386 (light grey) shows a signal at 4.4 days.
      At lower frequencies, the ground based photometry shows power and aliasing signals.
      The TESS signal shows a significant signal at 4.43 days, and a similar signal is seen in the ground-based data.
      The longer time baseline in the ASAS-SN data reveals substructure in the signal.}
        \label{fig:TESS_lc}
        \script{plot_tess_asas_ls_epochs.py}
    \end{centering}
\end{figure}

TESS data from the Quick-Look Pipeline \cite{2020RNAAS...4..204H,2021RNAAS...5..234K} was retrieved from the MAST archive and is shown in Extended Data Fig. \ref{fig:TESS_lc}.
The star was observed in sectors 8, 34 and 35.
The star was observed soon after the infrared brightening, and two sequential sectors some time later.

A periodic signal with a period of approximately 4.3 days is seen with a peak to peak amplitude of 2\% of the mean flux, and the similar period can be seen in the two later sectors but they are overwhelmed with the first signs of debris from the transiting object.
We carry out a Lomb-Scargle periodogram on S08, and obtain a significant detection of $P=4.43\pm 0.33$ days.
A similar period is seen at lower significance in the later sectors.
Aside from a peak at 1\,day, the  strongest peak in the ASAS-SN data has a similar period, at 4.1\,days.
We attribute this modulation due to star spots on the star rotating in and out of our view, and so we assert that is the rotational period of the star.
Using this rotational period the gyrochronological age \cite{Bouma23,Kounkel22} is calculated to be $300\pm92$ Myr.

\subsubsection*{Duration of the eclipse and gradient analysis}

\begin{figure*}
\begin{centering}
\includegraphics[width=1.0\textwidth]{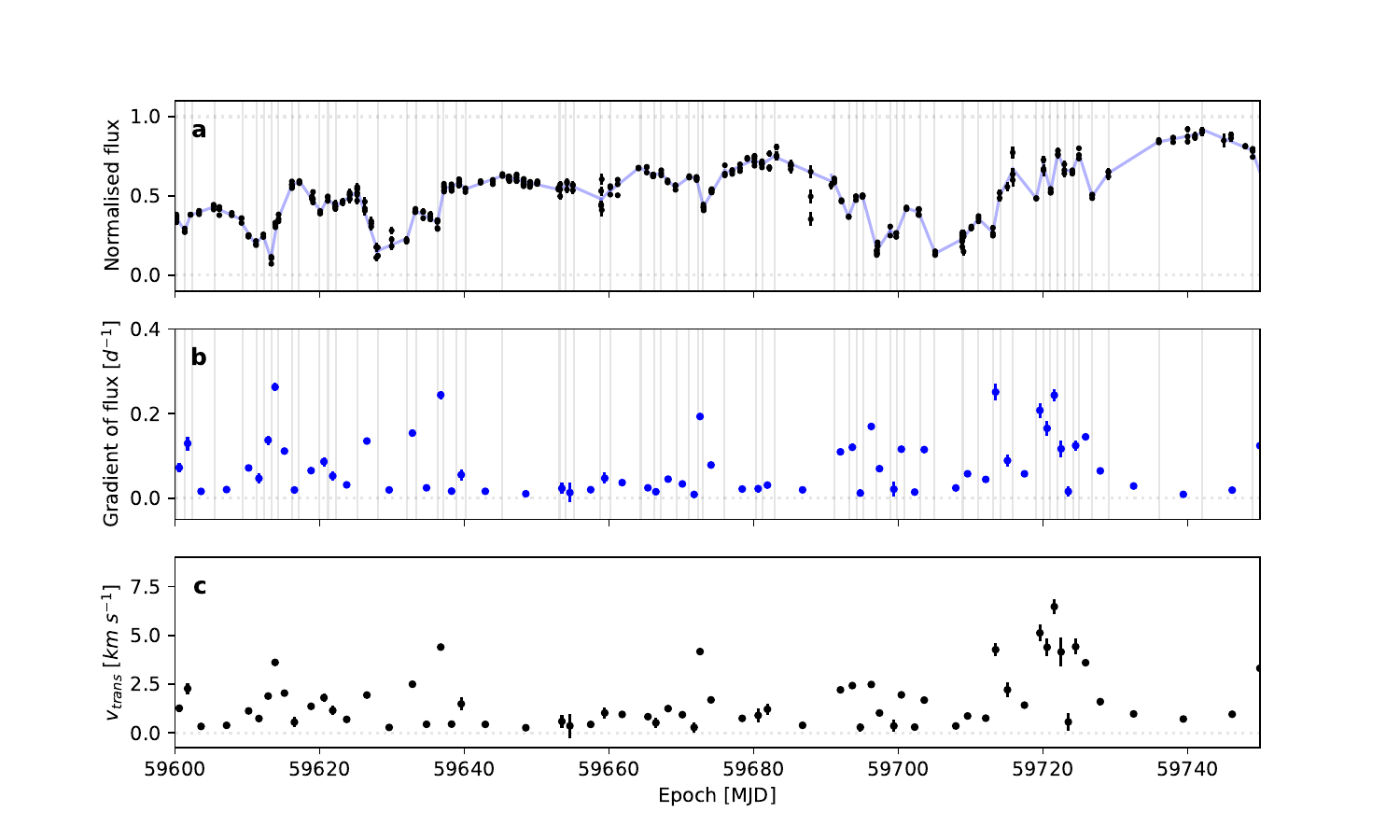}
      \caption{\textbf{Deriving the transverse velocity from a light curve.}
      {\bf a},ASAS-SN $g'$ photometry is shown in units of normalised flux.
      Straight line fits (light-blue lines) are made to the photometry in the regions indicated by the light-grey vertical lines.
      {\bf b},Gradient of the light curve as a function of time.
      {\bf c},Transverse velocity derived from the light curve and the gradient of the light curve.
      Error bars are shown at $1\sigma$ confidence.
}
        \label{fig:gradientconvert}
\script{plot_master_lightcurve_nature_with_zoom_3_panel.py}

\end{centering}
\end{figure*}

Figure \ref{fig:eclipse_overview} shows the optical light curves.
The start of the optical eclipse is seen around MJD 59350 in the $g'$ band observations, and returns to pre-eclipse levels by MJD 59850, giving a total eclipse duration of approximately 500 days.
The eclipse depth varies as a function of wavelength, which is discussed below.
The normalisation of the light curves for the ASAS-SN, ATLAS and ALLWISE photometry was by calculating the out of transit flux before MJD 58700.
The LCOGT and AAVSO photometry was determined by aligning the photometry of the eclipse with the ASAS-SN and ATLAS photometry on a per band basis.
By treating the linear changes in flux as due to the edges of large dust clouds crossing the disk of the star, and the largest amount of absorption within any of these segments as an estimate for the absorption of the cloud \citep[see equations 4.2 and 4.3 in ][]{Kennedy17}, we can determine a robust lower limit to the transverse velocity of 7.5 km\,s$^{-1}$ for the material moving in front of the star.
The method is illustrated in Extended Data Figure~\ref{fig:gradientconvert} and the results over the full optical transit are shown in the lower panel of Figure~\ref{fig:eclipse_overview}.
We convert the magnitude $M(t)$ at time $t$ to a normalised flux $f(t)=10^{\left(\left(M(t)-M_0\right)/-2.5\right)}$ where $M_0$ is the mean magnitude outside of the eclipse.
We visually determine turning points in the linearly increasing or decreasing photometric flux, fit straight lines to the selected points, and determine the flux gradient in units of day$^{-1}$.
A lower bound can be derived for the transverse velocity of the dust, $v$, by measuring the gradient of the light curve and determining what velocity a sharp edged and completely opaque occulter moving across the disk of the star would need to make the same gradient.
If the dust is on a circular orbit, it therefore has to be within 16\,au (equivalent to an orbital period of 63 years) around the star. The temperature of sub-micron dust grains at this distance would be in the range 100-200\,K.

\subsubsection*{Dust properties from optical colors}

\begin{figure}
    \centering
\includegraphics[width=1\textwidth]{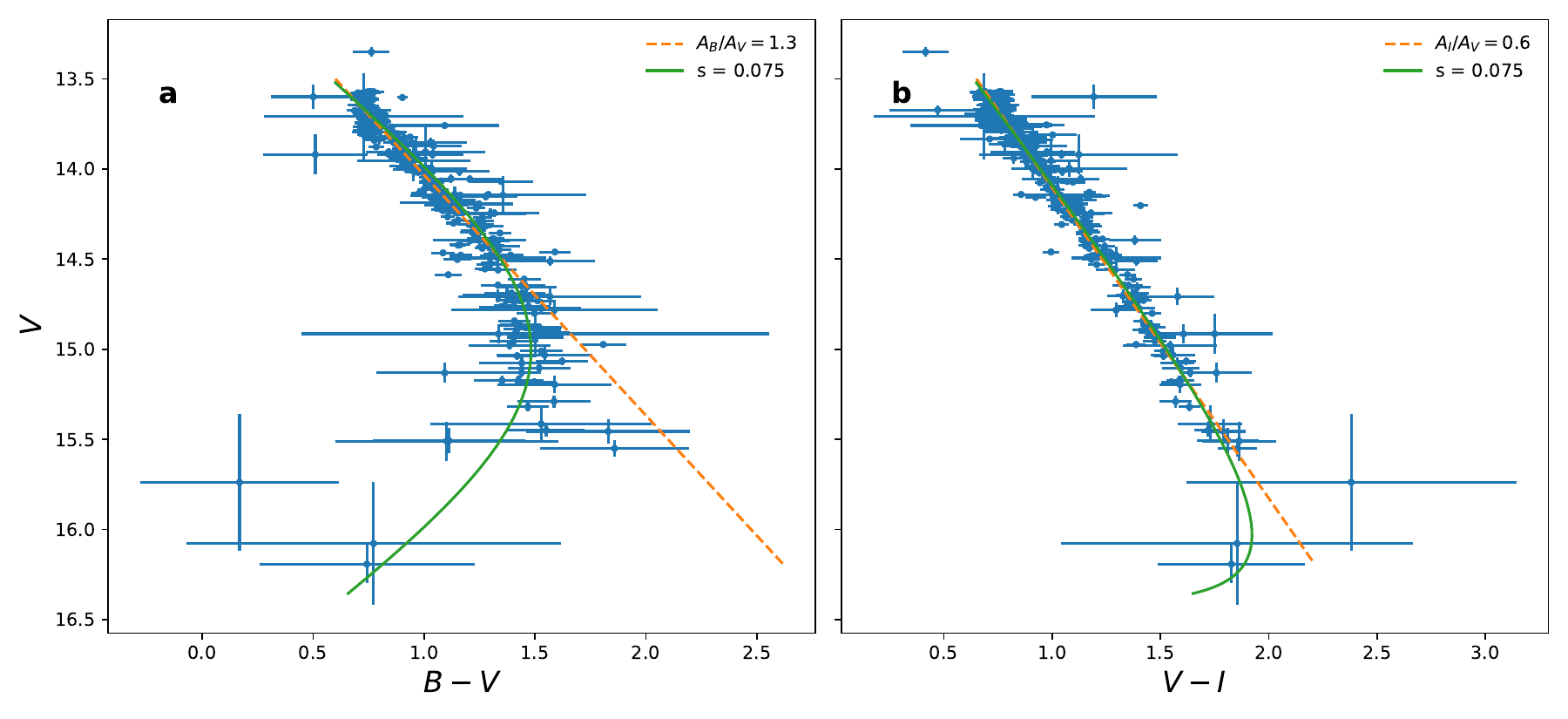}
    \caption{\textbf{Blueing of the $B-V$ color during dimming.}
    Points show AAVSO data, and lines show models.
    {\bf a},The $V$ magnitude versus $B-V$ color and {\bf b}, the  $V$ magnitude versus $V-I$ color.
    The dashed line is a line of $A_\lambda/A_V$ for the value shown in the legend and the solid line is a model that includes an underlying scattered-light component with $s=7.5$\% of the stellar flux.
    Error bars are shown at $1\sigma$ confidence.
}
    \label{fig:blueing}
    \script{blueing.py}
\end{figure}

As is clear from Figure \ref{fig:eclipse_overview} the photometry shows deeper absorption at bluer wavelengths compared to redder wavelengths.
This wavelength dependent absorption is typical of extinction due to particles with a characteristic size similar to or smaller than that of the observed wavelengths.
The differences are quantified in Extended Data Figure \ref{fig:blueing}, which shows the AAVSO $BVI$ photometric colours as a function of $V$ magnitude.
For both $B-V$ and $V-I$ the colour becomes redder as the star dims, and the reddening is here quantified by the total to selective extinction ratio $A_\lambda/A_V$ (dashed lines).
The values are similar to those seen for interstellar extinction \cite{1989ApJ...345..245C}, indicating that the dimming is caused by sub-micron sized dust \cite{1977ApJ...217..425M,2001ApJ...548..296W}.
A further constraint from the colours relates to scattering; in protoplanetary disk systems that undergo dimming (e.g. UX~Ori types) the colour initially reddens with dimming, but moves back towards the stellar colour when the dimming is more than one or two magnitudes \cite[e.g.][]{1994AJ....108.1906H}.
This ``blueing'' is typically interpreted as the relative increase in dust-scattered starlight from the disk or envelope as the star itself fades \cite{1988SvAL...14...27G}.
This behaviour is also seen in Extended Data Figure \ref{fig:blueing}, where $B-V$ shows significant blueing, while $V-I$ does not.
The solid lines show an extinction model where an underlying scattered light component with the same colour as the star has been added; as the star dims it reddens, but will eventually return to the stellar colour.
This happens more quickly for $B-V$ because nearly all of the stellar flux in $B$ is blocked, and is less pronounced for $V-I$ because the star is significantly less dimmed in $I$.
The fraction of scattered light is relatively large at 7.5\%, implying a significant complex of small dust around the star by the time the deepest parts of the optical transit occur.
This high fraction suggests that the impact occurred a significant fraction of an orbit before the optical transit, thus allowing the dust complex time to spread around the star.

\subsection*{Dust mass estimates}

The SED in Extended Data Figure \ref{fig:sed} gives an estimate of the infrared flux that would arise from dust thrown off in the putative collision.
We assume that the collision occurred at of order 10\,au from the star, and hence that a dust temperature of 100\,K is reasonable.
Here 1\% of the mass of a \,$\mathrm{M_{Earth}}$ planet is assumed to be converted entirely into 0.1\,$\mu$m sized grains; this is an optimistic assumption given that the total mass thrown off in collisions is of order 1\% or less \citep{2012ApJ...745...79L}, and that this ejected mass is constituted of bodies of a range of sizes and all the mass is not present as dust.
The dust spectrum in Extended Data Figure \ref{fig:sed} is approximated as a blackbody multiplied by $0.01 \times (10 / \lambda)^{1.5}$ (with $\lambda$ in $\mu$m), based on the absorption efficiency for silicates \citep{1993ApJ...402..441L}.
Changing the grain size to 1\,$\mu$m yields a similar dust spectrum; while the dust emitting area is less, these grains emit more efficiently.
Thus, the mass in small grains detectable with ALMA is about 0.2\,$M_{\rm Earth}$.
The dust spectrum lies well below the WISE measurements, and just below the ALMA measurement.
Even in this optimistic case, thermal emission from dust thrown off in the collision is therefore not necessarily easily detected.
This difference is one of surface area; at 600\,pc thousands of square au of 100\,K dust is needed for a thermal detection with ALMA, but only a small fraction of a square au is needed to significantly dim a star (which has radius $\sim$0.005\,au).

\subsection*{Alternative explanations for the observations}

\begin{figure}
    \centering
\includegraphics[width=1\textwidth]{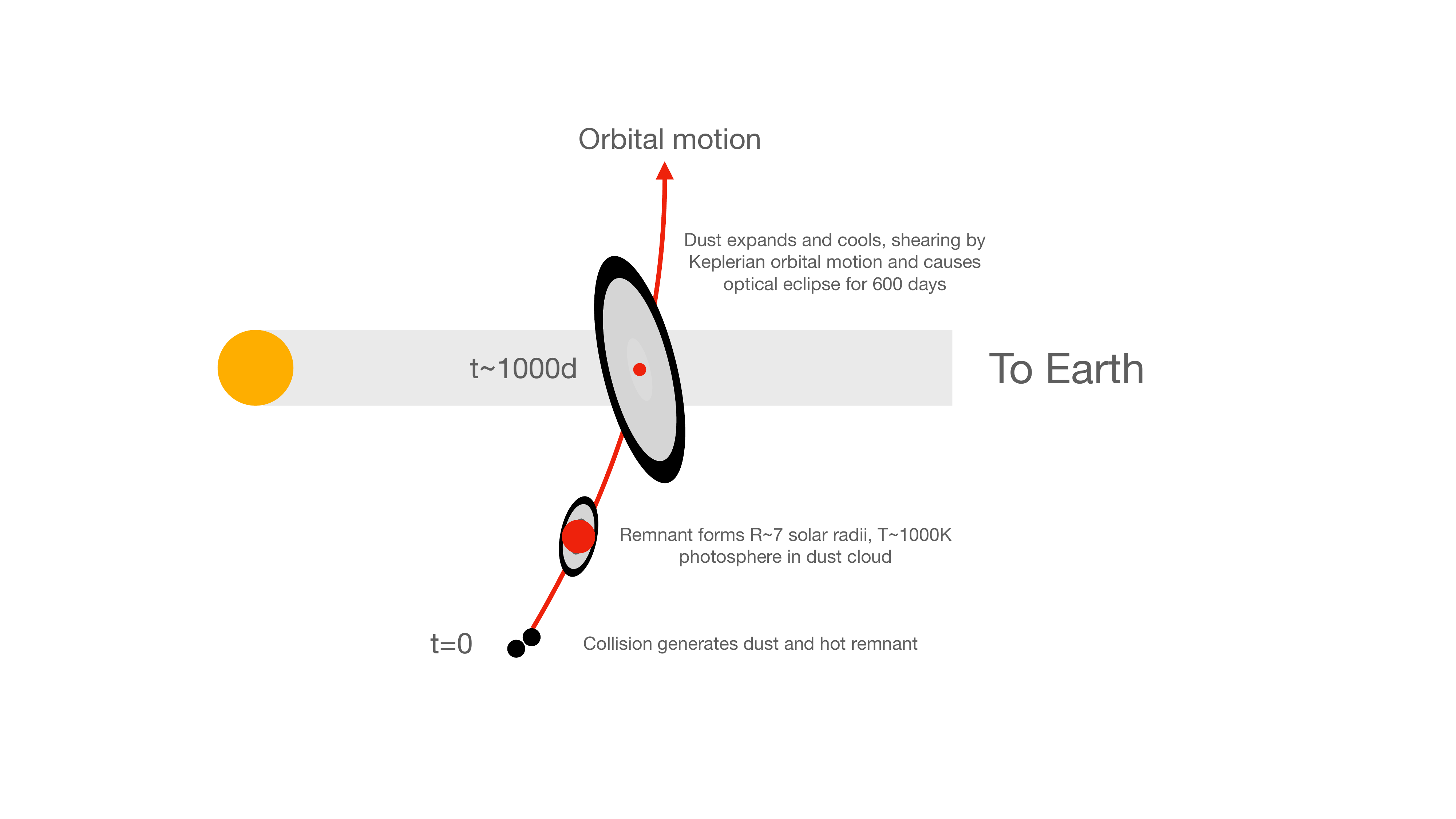}
    \caption{\textbf{Sketch of the hypothesis for the observations seen towards ASASSN-21qj.}
    At $t=0$, the collision occurs, producing a cloud of debris that expands and cools.
    Material close to the remnant is heated by its luminosity, generating the 1000\,K infrared emission.
    Around 1000 days later, the expanding cloud crosses the line of sight between the star and the Earth, generating the optical light curve.}
    \label{fig:hypothesis}
\end{figure}

We consider three possible scenarios to explain the observations: 1) The initial brightening and later eclipse are two unrelated phenomena.
2) The infrared emission and optical transit are both produced by a debris disk at $\sim0.1$~au.
3) We are observing the aftermath of a collision between super-Earths or mini Neptunes at at a semi-major axis of several au (see Extended Data Figure~\ref{fig:hypothesis}).
We provide here some more detail on the failures of the first two, and a preliminary simulation illustrating how an impact between two large bodies can produce a large object as hypothesised for our preferred scenario

First, the IR flux increase and the optical dimming might be unrelated, coincidental phenomena. 
For example, one debris disk at 0.1~au passively heated to 1000\,K producing the IR emission and another disk further from the star that transited \asas.
This explanation is unsatisfactory because both IR flux increases and dimming events are rare.
Mid-IR excesses are exceptionally rare among main-sequence stars \cite[1:10,000,][]{2013MNRAS.433.2334K}, and still uncommon for young stars \cite[1:100,][]{2013MNRAS.433.2334K}, and no star has previously shown a sizeable increase starting from no excess.
A single case of a disappearing mid-IR excess has been seen, which remains largely unexplained \cite{2012Natur.487...74M}.
Similarly, optical dimming events are rare for main-sequence stars, for example only one was seen to undergo dust-related optical dimming with Kepler main mission, which observed 150,000 stars for four years \cite{2016MNRAS.457.3988B}.
Both optical and IR variability are independently less than 1\% probabilities for a given star, so for \asas~to show both by chance is at best a 0.01\% probability, and probably much lower.

Another possibility is that both the IR and optical features are produced by a single debris disk at about 0.1\,au from the star.
At such a close distance, any dust clumps would initially produce periodic eclipses on the timescale of days \citep{2019MNRAS.488.4465G} before being sheared into an azimuthally symmetric structure in months, so the occultation of the star would need to be be related to changes in the vertical structure of a near edge-on post-collision disk, for example by dynamical ``stirring'' of debris by impact remnants \cite{1992Icar...96..107I}.
The optical depth of the disk must then decrease due to ongoing collisional depletion to explain the slow return to pre-transit levels of optical flux and gradual decrease in IR flux.
Three issues with this model are that 1) the disk must have precisely the right geometry to slowly occult the star as the scale height increased, and must coincidentally become optically thin, 2) there is no apparent change in dust temperature, which would be expected as the optical depth decreases and the warmer inner disk becomes visible, and 3) significant optical variation is seen three years after the putative collision, so any initially created clumps would already have sheared out.
Newer clumps must contribute of order 50\% of the dust area to explain the large variations around 59500\,MJD, but the IR flux shows a gradual decline rather than any strong variation that would be associated with clump creation.
A fourth, but less critical issue, is that the inferred clump velocities are not as high as they could be for transiting structures at 0.1\,au, which should result in transverse velocities of up to 100\,km\,s$^{-1}$.

\subsection*{SPH collision simulations}

\begin{figure}
    \centering
\includegraphics[width=1.0\textwidth]{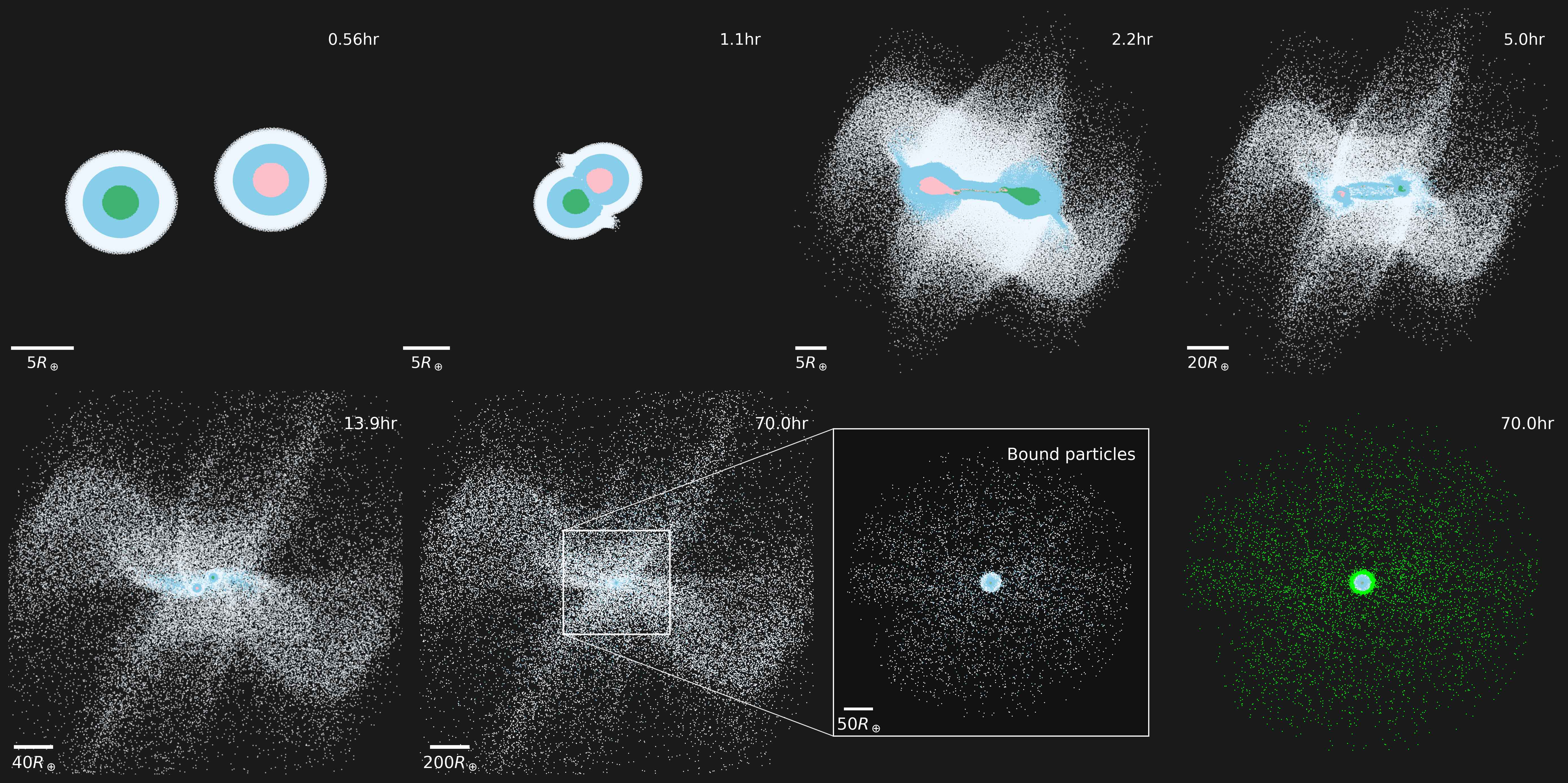}
    \caption{\textbf{Simulations of the formation of a post-impact body.}
    Giant impacts between super-Earths and mini-Neptunes can produce post-impact bodies hundreds of Earth-radii across, comparable with that required to produce the observed infrared flux.
    With the exception of the lower right panel, particles are colored by their material (forsterite, water, or a H$_2$-He mixture moving outwards in the initial bodies) and whether they came from the impactor or target (see top left panel).
    The final two panels show just the mass bound to the primary remnant which has a mass of 48.4~$M_{\rm Earth}$.
    In the final panel, particles that are at the minimum density imposed by the code are colored in green.}
    \label{fig:SPH}
\end{figure}

To provide some more insight into the collision scenario, we performed impact simulations using the SWIFT smoothed particle hydrodynamics (SPH) code \cite{Schaller2016,Schaller2018,Kegerreis2019}.
Extended Data Figure \ref{fig:SPH} shows a collision between two 25~$M_{\rm Earth}$ (Earth-mass) planets at 45.77~km~s$^{-1}$ (1.4~$v_{\rm esc}$, escape velocity neglecting the atmosphere) at an impact parameter of 0.4 (an impact angle of 23.6$^\circ$).
The colliding bodies were 22.5\% rock \cite[forsterite,][]{Stewart2019forsteriteEOS,Stewart2020_key_req_EOS}, 67.5\% water \cite{Senft2008}, and 10\% H/He \cite{Hubbard1980} by mass.
2.1$\times10^6$ particles were used in the simulation.
To make simulations with high resolution numerically tractable, SWIFT imposes a maximum smoothing length that, in effect, imposes a minimum density for particles in the simulation ($\sim30$~kg~m$^{-3}$ for the simulation shown here). 
The bound post-impact material is spread over hundreds of Earth radii following the collision, illustrating that giant impacts can produce very large post-impact objects. 

However, for the massive and highly extended bodies produced by collisions between super-Earths and mini-Neptunes, a large fraction of the post-impact body is at the minimum density (green particles in lower right panel in Extended Data Figure~\ref{fig:SPH}).
SPH simulations therefore likely underestimate the extent of such post-impact bodies, and further work is needed to fully quantify the size of post-impact bodies produced in different impacts.

\subsection*{Post-impact body cooling calculations}

How the emission from a post-impact body would evolve with time is highly dependent on the initial mass distribution and thermal state of the body, and the balance between radiative cooling, viscous spreading, and mass and angular momentum transport by condensates \cite{Lock2017,Lock18,Lock2020}. 
Given the limitations of SPH simulations (see above) it is not possible to accurately determine the initial structure of post-impact bodies in the relevant regime.
To explore a range of possible evolution pathways, we have calculated the evolution of post-impact bodies with different power-law surface density profiles, $\Sigma$, under the limiting case that radiative cooling and condensation of the vapor is the single driver for evolution of the structure. 
We chose a power-law surface density profile as it can straightforwardly cover the wide range of surface density profiles expected after super-Earth/mini-Neptune collisions, based on that found in impact simulations between lower-mass, terrestrial bodies \cite{Lock2017,Canup2001,Canup2012,Cuk2012,Rufu2017,Reufer2012}. 
The surface density profile is given by:
\begin{equation}
\Sigma=\begin{cases}
\Sigma_0, & \text{if } r_{xy}\leq R_{\rm c}\\
\alpha r_{xy}^{\beta}, & \text{if } R_{\rm c} < r_{xy}\leq R_{\rm emit}\\
0, & \text{if } r_{xy}> R_{\rm emit}
\end{cases}   
\label{eqn:sigma}
\end{equation}
where: $r_{xy}$ is the distance from the rotation axis; $R_{\rm c}$ is the outer radius of a constant surface density central region, roughly analogous to the corotating regions seen in Earth-mass synestias \cite{Lock2017,Lock18}; $\Sigma_0$ is the surface density of the central region; $\beta$ is the power law exponent; and $R_{\rm emit}$ is the initial emitting radius. Imposing surface density continuity gives
\begin{equation}
    \alpha=\Sigma_0 R_{\rm c}^{-\beta} \; .
\end{equation}
We can determine $\Sigma_0$ by fixing the mass of the body, 
\begin{equation}
    M_{\rm p} = \int_0^{R_{\rm c}}{2 \pi r'_{xy} \Sigma_0 dr'_{xy}} + \int_{R_{\rm c}}^{R_{\rm emit}}{2 \pi r'_{xy} \Sigma(r'_{xy}) dr'_{xy}} \; ,
\end{equation}
which can be solved to give:
\begin{equation}
\Sigma_0=\begin{cases}
\frac{M_{\rm p}}{\pi \left [R_{\rm c}^2 + 2 R_{\rm c}^{-\beta} \left ( \ln{R_{\rm emit}}-\ln{R_{\rm c}}\right ) \right ]}, & \text{if } \beta=-2,\\
\frac{M_{\rm p}}{\pi \left [R_{\rm c}^2 + \frac{2 R_{\rm c}^{-\beta}}{\beta+2} \left ( R_{\rm emit}^{\beta+2}-R_{\rm c}^{\beta+2}\right ) \right ]}, & \text{otherwise.} 
\end{cases}
\end{equation}
The time taken for a given region of the structure to cool to the point that a sufficient fraction of material is condensed for the temperature to drop below the condensation-buffer (see above) and so the emitted flux to drop is given by
\begin{equation}
    t_{\rm cool} (r_{xy})=\frac{l f(r_{xy}) \Sigma(r_{xy})}{\sigma T^4_{\rm emit}}
\end{equation}
where $f$ is the initial vapor fraction at that radius, $l$ is the latent heat of vaporization of the material (here we have taken the limiting case of pure water $l=2.256\times 10^6$ \cite{Chase1998}, but the addition of silicates could make the latent heat much larger), $\sigma$ is the Stefan–Boltzmann constant, and $T_{\rm emit}$ is the emission temperature.
Figure~\ref{fig:Hill_Bondi_R} shows example evolutions of the emission from a post-impact body using this model for different parameters (solid lines).
Results of a modified model where the initial surface density (by the addition of a constant parameter to Equation~\ref{eqn:sigma}) is forced to be zero at $R_{\rm emit}$ are shown as dashed lines.

\bmhead{Acknowledgments}

G.M.K is supported by the Royal Society as a Royal Society University Research Fellow.
S. J. L. acknowledges funding from the UK Natural Environment Research Council (grant NE/V014129/1).
L.C. acknowledges funding from the European Union H2020-MSCA-ITN-2019 under Grant Agreement no. 860470 (CHAMELEON)
J. D. acknowledges funding support from the Chinese Scholarship Council (No. 202008610218).
Giant impact simulations were carried out using the Isambard 2 UK National Tier-2 HPC Service (http://gw4.ac.uk/isambard/) operated by GW4 and the UK Met Office, and funded by EPSRC (EP/T022078/1).
We thank Krzysztof Stanek and the work of the ASAS-SN team with their survey and for providing public access to the database.
Part of this research was carried out at the Jet Propulsion Laboratory, California Institute of Technology, under a contract with the National Aeronautics and Space Administration (80NM0018D0004).

\bmhead{Authors' contributions}
MAK led the writing of the paper, management, obtaining the optical observations, and initial models.
SL led the afterglow modelling and theory.
GMK led the orbital analysis and dust analysis.
RvC did the optical light curve data reduction and reddening analysis and velocity constraint analysis.
EM performed the analysis of the properties of the star.
F.-J.H and EG carried out optical monitoring of the star.
JM, AM, JDK and AS were responsible for NEOWISE identification and data reduction.
SL, LC, JD, PT and ZL provided discussion on the ejected material and subsequent evolution.
JD performed the SPH impact simulations.
HB, SC, OG, PLD, LM and PT were responsible for the observation and reduction of observational data.
MRS led the discovery of the optical dimming of the star.
All co-authors assisted with manuscript writing and proofreading.

\bmhead{Competing interests} The authors declare no competing interests.

\bmhead{Availability of data and materials}
The data sets generated and analysed during the current study are available in the Zenodo repository \url{https://doi.org/10.5281/zenodo.8344755}.

\bmhead{Code availability} 
All the code for the analysis and the generation of all the figures  are available in a showyourwork \citep{Luger2021} reproducible framework available as a git repository on github: \url{https://github.com/mkenworthy/ASASSN-21qj-collision/}.
The source code and documentation for the SWIFT open-source simulation code is available from \url{www.swiftsim.com}.

\bmhead{Correspondence and requests for materials}
should be addressed to Matthew Kenworthy.

\bmhead{Reprints and permissions information} is available at http://www.nature.com/reprints.
\newpage

\bibliography{bib}

\end{document}